\documentclass[conference,letterpaper]{IEEEtran}
\IEEEoverridecommandlockouts
\usepackage{cite}
\usepackage{amsmath,amssymb,amsfonts}
\usepackage{algorithmic}
\usepackage{graphicx}
\usepackage{textcomp}
\usepackage{import}
\usepackage{placeins}
\usepackage{textcomp}
\usepackage{url}
\usepackage[hypertexnames=false,bookmarks=false]{hyperref} 
\hypersetup{colorlinks,linkcolor={red},citecolor={blue},urlcolor={black}} 
\usepackage[euler]{textgreek} 
\usepackage{blindtext}
\usepackage{ragged2e}
\usepackage[table,xcdraw]{xcolor}
\usepackage{tabularx}
\usepackage{multirow} 
\usepackage{comment}
\usepackage{stfloats}

\newcolumntype{s}{>{\hsize=.5\hsize}X}
\newcolumntype{q}{>{\hsize=.25\hsize}X}

\ifCLASSOPTIONcompsoc
    \usepackage[caption=false, font=normalsize, labelfont=sf, textfont=sf]{subfig}
\else
\usepackage[caption=false, font=footnotesize]{subfig}
\fi

\def\BibTeX{{\rm B\kern-.05em{\sc i\kern-.025em b}\kern-.08em
    T\kern-.1667em\lower.7ex\hbox{E}\kern-.125emX}}

\IEEEaftertitletext{\vspace{-0.7\baselineskip}}
    
\begin{document}
\bstctlcite{IEEEexample:BSTcontrol}

\title{A mixed-signal analogue front-end for brain-implantable neural interfaces using a digital fixed-point IIR filter and bulk offset cancellation
\vspace{-8mm}

}

\author{%
\IEEEauthorblockN{Dimitris Antoniadis$^{1,2}$, Timothy G. Constandinou$^{1,2,3}$}
\IEEEauthorblockA{$^{1}$Dept. of Electrical \& Electronic Engineering, Imperial College London, South Kensington Campus, London, SW7 2AZ, UK}
\IEEEauthorblockA{$^{2}$UK Dementia Research Institute Centre for Care Research \& Technology at Imperial College London \& University of Surrey}
\IEEEauthorblockA{$^{3}$Mint Neurotechnologies Ltd, 125 Wood Street, London, EC2V 7AW, UK}
\IEEEauthorblockA{Email: \{dimitris.antoniadis20, t.constandinou\}@imperial.ac.uk}
}

\IEEEaftertitletext{\vspace{-8mm}}
\maketitle

\begin{abstract}

Advances in miniaturised implantable neural electronics have paved the way for therapeutic brain–computer interfaces with clinical potential for movement disorders, epilepsy, and broader neurological applications. This paper presents a mixed-signal analogue front end (AFE) designed to record simultaneously both extracellular action potentials (EAPs) and local field potentials (LFPs). The feedforward path integrates a low-noise amplifier (LNA) and a successive-approximation-register (SAR) analogue-to-digital converter (ADC), while the feedback path employs a fixed-point infinite-impulse-response (IIR) Chebyshev Type II low-pass filter to suppress sub-mHz components via bulk-voltage control of the LNA input differential pair using two R–2R pseudo-resistor digital-to-analogue converters (DACs). The proposed AFE employs a low-power (LP) mode and an offset-cancellation high-performance (HP) mode. 
The proposed AFE achieves 40.55\,dB gain and supports neural recording from 0.1\,Hz to 5.705\,/\,9.66\,kHz\,(LP\,/\,HP), with typical input-referred noise of 3.9\,/\,6.615\,\textmu V\textsubscript{rms} in the LFP band and 11.42\,/\,11.11\,\textmu V\textsubscript{rms} in the EAP band (LP\,/\,HP). Its typical power per channel is 5.44\,\textmu W (LP) and 11.35\,\textmu W (HP), while it occupies 0.198 mm\textsuperscript{2}.
\end{abstract}
\section{Introduction} \label{sec:Introduction}

Advances in miniaturised implantable neural electronics have enabled
brain-machine interface applications~\cite{rapeaux2021implantable, ahmadi2019towards, muller2021miniaturized, Szostak2020, musk2019integrated},
with growing clinical potential for movement disorders, epilepsy, and
broader neurological and bioelectronic-medicine applications~\cite{wolpaw2020brain, brunner2015bnci, mcfarland2008emulation, robinson2021emerging, gonzalez2024bioelectronic, lerman2025next}.
A key requirement for implantable neural interfaces is detecting extracellular action potentials (EAPs), typically 50\,\textmu V\textsubscript{pp} to 500\,\textmu V\textsubscript{pp} from 100\,Hz to 10\,kHz~\cite{kandel2000principles, mollazadeh2009wireless}, and sensing local field potentials (LFPs) of 0.5\,mV\textsubscript{pp} to 5\,mV\textsubscript{pp} across 100\,mHz to 200\,Hz \cite{gosselin2011recent, white2010real, patil2008development}. 


The analogue front end (AFE) of a neural implant handles neural signal acquisition, typically consisting of low-noise amplifiers (LNAs) followed by multiplexing into an analogue-to-digital converter (ADC). 
Miniaturised electrodes at the tissue interface introduce large,
time-varying DC offsets~\cite{woeppel2017recent,villa2024enhancing,campbell2018chronically}. AC-coupled AFEs reject
these offsets and provide electrical isolation~\cite{uran2020ac,bagheri2016low}, but
sub-hertz high-pass behaviour demands prohibitively large coupling
capacitors, limiting scalability in high-channel-count
systems~\cite{uran2020ac}.
One DC-coupled approach achieves offset tolerance through bulk-voltage control via auxiliary analogue circuitry~\cite{sporer2022direct}, though the high pass corner is implemented using pseudoresistors and capacitors. Mixed-signal techniques including direct digitisation, delta-modulation, switched-capacitor feedback, and input multiplexing have been explored, yet often at the expense of input impedance or increased switching noise~\cite{kassiri2017rail,uehlin20190,de2018fully}. Recent methods employ digital low-frequency feedback for offset suppression, however the digital filter has been implemented off chip~\cite{muller20110}. Another approach implements the digital low pass filtering on chip, achieving better scalability but captures only LFPs~\cite{bagheri2016low}. Additionally, a common issue neglected on both DC and AC-coupled neural implants is that one-time post-fabrication digital trimming schemes used for performance configuration are ineffective due to packaging limitations.

\begin{figure*}
    \centering
    \includegraphics[width=\textwidth]{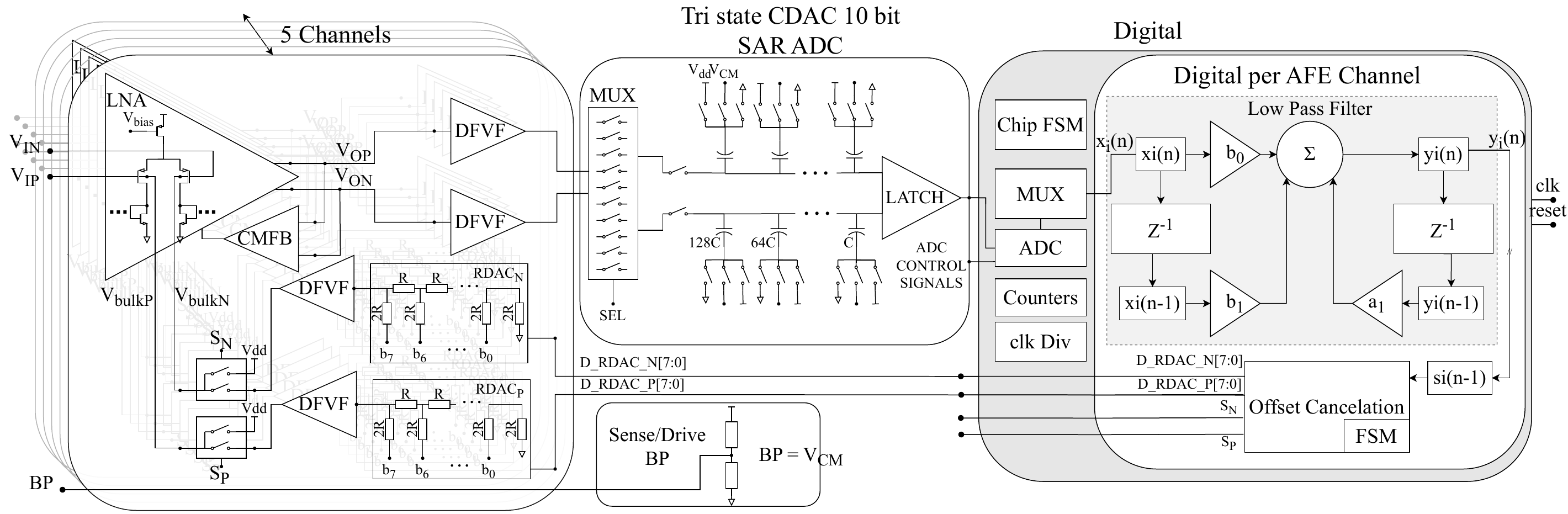}
    \caption{Architecture of the proposed AFE.}
    \label{fig:AFE_architecture}
    
\end{figure*}

This paper presents a mixed-signal DC-coupled AFE for implantable neural recording that combines and enhances previously established offset-cancellation techniques in three ways. First, body potential (BP) referenced biasing is used for all circuits, eliminating reliance on post-fabrication digital trimming to compensate for process-voltage-temperature (PVT) variations — a practical necessity given packaging constraints in implantable systems. Second, building on the principle from \cite{sporer2022direct}, electrode DC offset suppression is achieved via bulk-voltage control of the LNA input differential pair, driven by two R–2R pseudo-resistor DACs with a digital servo-path extending the approach from \cite{muller20110}. Third, the required sub-hertz low-pass filtering is realised on-chip using a fixed-point IIR filter in the feedback path — unlike prior work where this is implemented off-chip. An autonomous high-performance (HP) mode is additionally incorporated, activating the offset-cancellation loop upon detection of a significant input offset. The proposed architecture is illustrated in Fig.~\ref{fig:AFE_architecture}.

Section~\ref{sec:AFE_architecture} details the proposed AFE architecture, Section~\ref{sec:results} reports the simulation results, and Section~\ref{sec:conclusion} summarises the conclusions and outlines future work.

\section{AFE Architecture} \label{sec:AFE_architecture}
\subsection{AFE Architecture Overview}

Packaging constraints preclude the use of digital trimming techniques to compensate for PVT variations,  necessitating a design approach referenced to a known  on-chip potential. A convenient reference is the BP, which can be either sensed or driven depending on the grounding strategy~\cite{haci2018design}. In this work, BP is set to half the supply voltage and used as the common reference, with all behaviour defined relative to this level.

The system starts at a low-power mode (LP) and the differential neural signal is sensed at the electrode interface around the body potential (BP) common-mode reference. The signal is amplified by the low-noise amplifier (LNA) and buffered using two differential flipped-voltage followers (DFVFs). The buffered signal is then multiplexed to a successive-approximation-register (SAR) analogue-to-digital converter (ADC) incorporating a tri-level capacitive DAC (CDAC)~\cite{yuan2012low}. This enables the CDAC conversion to operate around the BP common-mode level. The digitised output is passed through a Chebyshev Type II sub-hertz low-pass filter, which extracts the low-frequency offset component used for digital offset-cancellation control. When this component exceeds a predefined threshold, the high-performance (HP) mode is activated, engaging the offset-cancellation loop. The bulk voltage of the LNA input differential pair is now driven by the outputs of two R–2R DACs, while the LNA bias current is increased to reduce input-referred noise. 
At reset both DACs are set to maximum code. On detecting a positive offset, RDAC\_P decrements, increasing current through the corresponding PMOS device, until the extracted low-frequency component re-enters the target range, at which point the loop holds until re-triggered.

\begin{table}[!t]
\centering
\caption{Post Layout LNA Performance in Low-Power Mode}
\label{tabla_lp_mode}
\begin{tabular}{lccccc}
\hline
\textbf{Metric} & \textbf{Min} & \textbf{Max} & \textbf{Mean} & \textbf{MC Mean} & \textbf{MC $\sigma$} \\
\hline
$I$ (\textmu A) & 1.37 & 2.553 & 1.758 & 1.672 & 0.15 \\
$f_c$ (kHz) & 3.92 & 8.26 & 5.705 & 5.55 & 0.61 \\
Gain (dB) & 38.26 & 40.92 & 39.7 & 39.66 & 0.565 \\
Noise$_\text{LF}$ (\textmu V\textsubscript{rms}) & 3.593 & 4.233 & 3.904 & 3.91 & 0.064 \\
Noise$_\text{HF}$ (\textmu V\textsubscript{rms}) & 9.786 & 13 & 11.42 & 11.47 & 0.416 \\
\hline
\multicolumn{6}{l}{\footnotesize Noise$_\text{LF}$: 0.1--200 Hz. Noise$_\text{HF}$: 200 Hz--10 kHz.} \\
\end{tabular}
\end{table}

\begin{table}[!t]
\centering
\caption{Post Layout LNA Performance in Offset-Cancellation Mode}
\label{tabla_hp_mode}
\begin{tabular}{lccccc}
\hline
\textbf{Metric} & \textbf{Min} & \textbf{Max} & \textbf{Mean} & \textbf{MC Mean} & \textbf{MC $\sigma$} \\
\hline
$I$ (\textmu A) & 5.253 & 8.71 & 6.69 & 6.571 & 0.503 \\
$f_c$ (kHz) & 7.245 & 12.97 & 9.66 & 9.427 & 1.408 \\
Gain (dB) & 39.46 & 41.4 & 40.55 & 40.39 & 0.685 \\
Noise$_\text{LF}$ (\textmu V\textsubscript{rms}) & 6.431 & 6.834 & 6.615 & 6.619 & 0.351 \\
Noise$_\text{HF}$ (\textmu V\textsubscript{rms}) & 9.42 & 12.76 & 11.11 & 11.15 & 0.37 \\
\hline
\multicolumn{6}{l}{\footnotesize Noise$_\text{LF}$: 0.1--200 Hz. Noise$_\text{HF}$: 200 Hz--10 kHz.} \\
\end{tabular}
\end{table}

The results presented in the following subsections are obtained from simulations across combinations of typical, fast, and slow corners, as well as Monte Carlo (MC) analysis including device process and mismatch variation, evaluated at temperatures of 25, 36, and 47\textdegree C. While body temperature is around 36\textdegree C, the two additional temperatures are used to stress the robustness of the devices.

\subsection{Low Noise Amplifier}

The technology supports a 1.2\,V power supply, so a conventional fully differential current mirror  LNA topology was selected to maximise the available dynamic range. 
With the BP set at half the supply voltage, the differential input pair always operates at an appropriate DC bias level. The common-mode feedback (CMFB) amplifier reference voltage is also set to BP, ensuring proper biasing throughout the signal path.
Tab.~\ref{tabla_lp_mode} and Tab.~\ref{tabla_hp_mode} report LP and HP performance when driving a DFVF and an additional conservative  load of 10\,m$\Omega$ and 300\,fF. The increased HP current arises from driving the LNA bulk terminals through the DFVF output devices.
Because the g\textsubscript{mb} of the input PMOS transistors is lower than their g\textsubscript{m}, noise at the DFVF output has significant impact when referred to the LNA input. To reduce this noise to acceptable levels, the LNA bias current is increased during offset cancellation.  
Analysis showed that 1\,mV of input  offset can be compensated by applying approximately 2.56\,mV to the bulk terminals of the differential input pair.
This relationship remains approximately linear over the intended offset-cancellation range.

\subsection{Differential Flipped Voltage Follower}

Two variants of a pseudo-differential flipped-voltage follower (DFVF) buffer are employed, chosen for their low power consumption and push–pull output stage, enabling wide bandwidth~\cite{ramirez2006new}. 
The LNA output is buffered using two DFVFs to drive the multiplexer and SAR ADC CDAC ($\approx$30\,pF). Additionally, the outputs of the R–2R DACs are buffered using ultra-low-power DFVFs, 
where the offset-cancellation loop handles only sub-hertz components, reducing bandwidth and current requirements.
The performance of the high-bandwidth DFVF driving a conservative 60\,pF load is summarised in Tab.~\ref{tab:DFVF} and the low-current DFVF when driving 5\,pF is shown in Tab.~\ref{tab:DFVF_DAC}. A positive input offset limits buffer accuracy but is negligible due to its symmetric appearance in the fully differential path.




\begin{table}[!tb]
\centering
\caption{Post Layout Performance of DFVF Buffer at LNA Output}
\label{tab:DFVF}
\begin{tabular}{lccccc}
\hline
\textbf{Metric} & \textbf{Min} & \textbf{Max} & \textbf{Mean} & \textbf{MC Mean} & \textbf{MC $\sigma$} \\
\hline
$I$ (\textmu A) & 0.635 & 0.907 & 0.747 & 0.747 & 0.07 \\
GBW (kHz) & 316.9 & 429.6 & 366.5 & 364.1 & 44.46 \\
Loop Phase Margin (°) & 53.33 & 62.45 & 57.97 & 57.32 & 3.42 \\
Error (mV) & 3.944 & 8.93 & 5.77 & 5.62 & 4.18 \\
\hline
\end{tabular}
\end{table}

\begin{table}[!tb]
\centering
\caption{Post layout Performance of DFVF at RDAC Output}
\label{tab:DFVF_DAC}
\begin{tabular}{lccccc}
\hline
\textbf{Metric} & \textbf{Min} & \textbf{Max} & \textbf{Mean} & \textbf{MC Mean} & \textbf{MC $\sigma$} \\
\hline
$I$ (\textmu A) & 0.132 & 0.222 & 0.158 &  0.151 & 0.034 \\
GBW (kHz) & 145.2 & 211 & 198 & 204 & 14.3 \\
Loop Phase Margin (°) & 36.2 & 39.1 & 37.8 & 36.5 & 0.63 \\
Error (mV) & 5.67 & 7.1 & 6.34 & 6.27 & 0.86 \\
\hline
\end{tabular}
\end{table}

\subsection{Resistor DAC}

In order to avoid excessive area, the 8-bit R–2R DAC in this design uses pseudo-resistors. Each R unit comprises four pseudo-resistors in series. 
Across all corners, the effective resistance of R unit ranges from 27.82\,T$\Omega$ to 136\,T$\Omega$. The R unit is also sensitive to RDAC code changes. During code sweep, it exhibits an average step size of 266\,\textmu V, with a minimum of 170\,\textmu V and a maximum of 2.96\,mV at specific transition codes over PVT. This allows the offset-cancellation algorithm to typically reduce the residual offset to around 100 \textmu V, with a worst-case code residual of approximately 1.15\,mV. In MC simulation, it exhibits mean step value of 262.5\,\textmu V ($\sigma$\,=\,6.83\,\textmu V). The MC mean of max step during code sweep is 2.298\,mV ($\sigma$\,=\,490\,\textmu V). It always remains monotonic. Worst case code residual does not saturate the dynamic input range of the LNA and the offset can later be removed in digital post processing. The RDAC output range spans from 750\,mV to 899\,mV, ensuring the bulks of the differential pair never become forward-biased. Depending on the code, the current consumption ranges from 1\,nA to 150\,nA and for MC from 2.5\,nA to 91.8\,nA (MC mean\,=\,13.78\,nA, $\sigma$\,=\,14.83\,nA).

\subsection{Tri-state CDAC SAR ADC}

The ADC is shared across multiple channels. The chosen architecture is based on the 10-bit differential SAR ADC described in \cite{yuan2012low}, which employs top-plate sampling on the capacitive DAC (CDAC). 
The MSB is set by the first comparison and the remaining bits are resolved by binary search, with the losing side referenced to  V\textsubscript{cm}\,=\,BP. Conversion thus remains centred on the BP level. 
Operating around V\textsubscript{cm}, the zero-offset reference of the low-pass filter, ensures consistent behaviour across the signal chain and
inherent common-mode rejection.
The ADC draws 4.12\,\textmu A per conversion on average, with 3.64 and 6.27\,\textmu A min and max respectively. The 9.3\,\textmu s conversion time yields a 107\,kHz maximum rate, sufficient
to sample five channels at 20\,kHz for signals below 10\,kHz.
The digital power consumption for ADC SAR controller is estimated equal to 2.748\,\textmu W from final gds implementation netlist.

\subsection{Digital Low Pass Filter}

The digital filter in the feedback path must be as compact and power-efficient as possible. To meet these constraints, an IIR architecture was selected, enabling a low-order implementation with reduced hardware complexity. To minimise area, a fixed-point design was adopted with the smallest coefficient and data word lengths that maintain accuracy. The filter uses a single biquad section—the basic second-order IIR building block—in a Direct Form~I structure for reduced hardware and improved overflow handling.
A Chebyshev Type~II low-pass filter was designed with a 16\,kHz 
sampling frequency, 1\,mHz passband edge, 0.1\,Hz stopband edge, 
0.01\,dB passband ripple, and 50\,dB stopband attenuation. 
Fixed-point precision was determined using an adapted methodology 
from~\cite{volkova2018towards}, evaluating the L1-norm difference 
between the ideal double-precision and fixed-point impulse responses 
over $2^{21}$ samples. The filter output requires 11 integer bits, 
and a ``sweet spot'' of 40 fractional bits was identified where the 
quantisation error approaches zero, with internal word widths sized 
accordingly. The digital power consumption is estimated equal to 230\,nW from final gds implementation netlist.

\begin{figure}[t]
    \centering\centerline{\includegraphics[width=0.99\columnwidth]{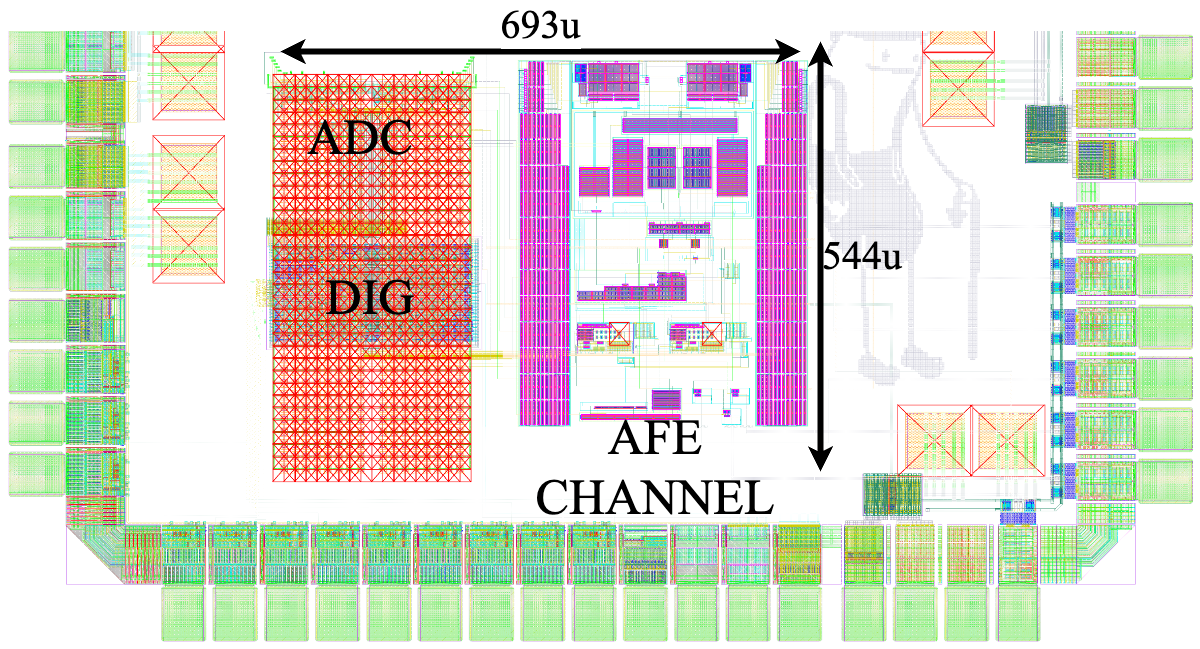}}
    \caption{AFE Layout.}
    \label{fig:AFE_layout}
\end{figure}
\section{Results} \label{sec:results}

The layout shown in Fig.~\ref{fig:AFE_layout}, being fabricated in 65\,nm CMOS, occupies
693\,\textmu m\,$\times$\,544\,\textmu m total: 467\,\textmu m\,$\times$\,376\,\textmu m
(0.17\,mm$^2$) for the LNA with buffers and 544\,\textmu m $\times$
258\,\textmu m (0.14\,mm$^2$) for the ADC. As the ADC area is mostly capacitors, the underlying region remains available for digital logic.

\begin{table*}[b]
\centering
\caption{Performance Comparison of Neural Recording AFEs}
\label{tab:afe_comparison}
\setlength{\tabcolsep}{5pt} 
\begin{tabular}{lccccc}
\hline
\textbf{Metric} & \textbf{\cite{muller20110}} & \textbf{\cite{sporer2022direct}} & \textbf{\cite{wendler202128}}  & \textbf{\cite{pochet2021174}} & \textbf{Typical This work\textsuperscript{$\dagger$}} \\
\hline
Technology (nm) & 65 & 180  & 180 & 65 & 65 \\
Supply (V) & 0.5 & 1.8 & 1.8  & 1.2/0.8 & 1.2 \\
Power/ch (\textmu W) & 5.04 & 12.8 & 14.94  & 5.8 & 5.44\textsuperscript{*} (LP), 11.35\textsuperscript{*} (HP) \\
Area/ch (mm$^2$) & 0.013 & 0.020 & 0.0046  & 0.075 & 0.198 \\
Bandwidth (kHz) & 10 & 10 & 10 & 1 & 5.705 (LP), 9.66 (HP) \\
Peak input (mV$_\text{pp}$) & -- & 19 & 14 & 400 & 16.8\textsuperscript{$\ddagger$} \\
EDO range (mV) & $\pm$50 & $\pm$100 & $\pm$60  & $\pm$200 & $\pm$58.125 \\
LFP noise (\textmu V$_\text{rms}$) & 4.3 @ 10--300 Hz & 1.82 @ 1--200 Hz & 2.72 @ 0.5--1000 Hz & 3.5 @ 1--1000 Hz & 3.904 (LP), 6.615 (HP) @ 0.1--200 Hz \\
EAP noise (\textmu V$_\text{rms}$)  & 4.9 @ 0.3--10kHz Hz & 5.55 @ 0.2--10 kHz & 4.37 @ 0.3--10 kHz & --  & 11.42 (LP), 11.11 (HP) @ 0.2--10 kHz \\

NEF  &  5.99 & 5.75 & 5.01 & --  & 9.91 (LP), 15.33 (HP) \\

PEF  & 17.96 & 59.5 & 45 & -- & 117.8 (LP), 282.3 (HP) \\
\hline
\multicolumn{6}{l}{EDO: electrode DC-offset cancellation range. LP/HP: low-power / offset-cancellation modes for this work. \textsuperscript{$\ddagger$}At
10\,\% gain deviation.} \\
\multicolumn{6}{l}{\textsuperscript{$\dagger$}This work presents simulation results, while \cite{muller20110}, \cite{sporer2022direct}, \cite{wendler202128}, \cite{pochet2021174} are silicon measured. \textsuperscript{*}Total digital power consumption per channel included. } \\
\end{tabular}
\end{table*}

\begin{figure}[t]
    \centering
    \centerline{\includegraphics[width=\columnwidth, trim=0pt 10pt 0pt 20pt, clip]{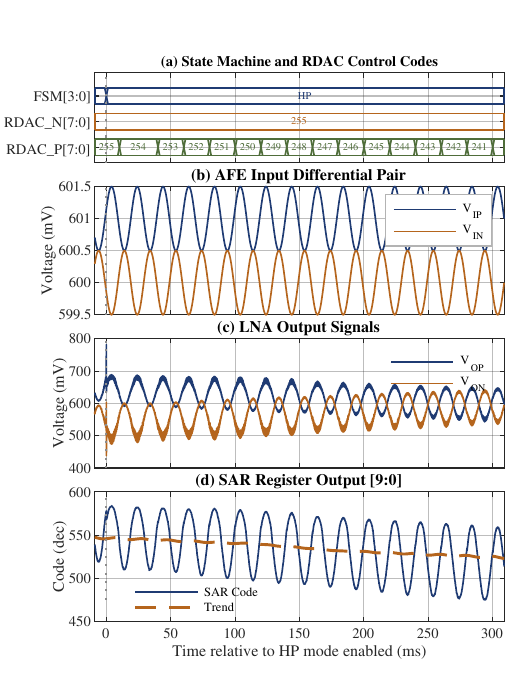} }
    \caption{AFE operation with a 1\,mV, 50\,Hz differential input and 1\,mV DC electrode offset, demonstrating the offset cancellation mechanism converging around mid code 512. For demonstration purposes a faster prescaler clock has been applied at RDAC transitions.}
    \label{fig:AFE_performance_no_offset}
\end{figure}

Fig.~\ref{fig:AFE_performance_no_offset} demonstrates the offset-cancellation mechanism for a 50\,Hz, 1\,mV input with 1\,mV DC offset, with the LNA output sampled at 19.53\,kHz. As shown in Fig.~\ref{fig:AFE_performance_no_offset}(a), once HP mode is enabled the RDAC\_P code decrements and the LNA output settles around the BP common-mode level, while the SAR code converges to mid-code 512 as depicted in Fig.~\ref{fig:AFE_performance_no_offset}(d).
The total current consumption per channel with digital included is typically 4.534\,\textmu A (LP) and 9.46\,\textmu A (HP).
Based on the dynamic range of the RDAC, the system can suppress low-frequency offset components up to $\pm$58.13\,mV. 
The offset-cancellation loop typically reduces the residual offset to
$\approx$100\,\textmu V (worst-case 1.15\,mV), a quasi-static DC term that does not interfere with detection and is removable in digital post-processing. The typical LNA linear input range is 16.8\,mV\textsubscript{pp} (4.8\,mV\textsubscript{pp} worst-case) at
10\,\% gain deviation. The channel
therefore keeps recording throughout cancellation, at full fidelity when
the offset is within range and with increasing distortion for larger
initial offsets.


\begin{figure}[t]
    \centering\centerline{\includegraphics[width=0.99\columnwidth]{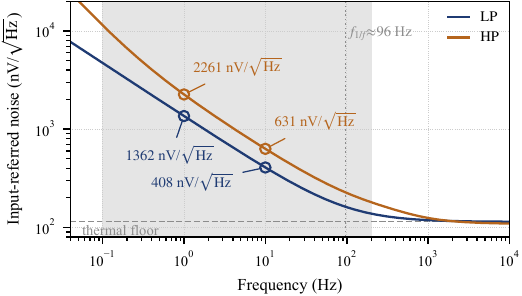}}
    \caption{Input-referred noise PSD in LP and HP modes (typical corner). The shaded region marks the 0.1\,–\,200\,Hz LFP integration band. The 1/f corner lies at $\approx$\,96\,/\,334\,Hz (LP\,/\,HP), above which the thermal floor ($\approx$\,115\,nV/$\sqrt{\text{Hz}}$) dominates. Spot values are annotated at 1 Hz and 10 Hz.}
    \label{fig:psd}
\end{figure}

A performance comparison with recent state-of-the-art AFEs, for both typical case LP and HP offset-cancellation modes, is provided in Tab.~\ref{tab:afe_comparison}. The proposed implementation achieves relatively good current and power consumption  with digital included for an increased footprint, while HP mode trades noise and efficiency for offset-rejection performance. At LP the proposed implementation does not cover the whole 10\,kHz range indicating further improvements could be explored.
Its NEF and PEF is higher than the compared works, as expected due to its higher noise. HP mode incurs the expected penalty of the additional RDAC and DFVF bulk-driven path, with PEF rising to 282.3, reflecting the g\textsubscript{mb}\,$<$\,g\textsubscript{m} penalty.
The larger area footprint is a consequence of using long-channel devices to maintain current mirror saturation and robustness across PVT.
The ADC input-referred LSB of $\approx$ 12\,\textmu V is comparable to the LNA noise floor, suggesting moderate noise-power trade-offs remain available for improvement in future iterations. 
Fig.~\ref{fig:psd} shows the typical input-referred noise power spectral density (PSD). The 1/$f$ corner is $\approx$\,96\,/\,334\,Hz (LP\,/\,HP), with spot values 1362\,nV/$\sqrt{\mathrm{Hz}}$\,(LP) at 1\,Hz and 631\,nV/$\sqrt{\mathrm{Hz}}$\,(HP) at 10\,Hz. 
Finally, a limitation is the long settling time of the sub-blocks, imposed by the sub-hertz feedback filter. The chip startup settling is set at 525\,s for worst case, the initial RDAC settling can reach 200\,s, and ${\approx}$\,33 s are required after each 1-bit RDAC update before the next. During the offset cancellation state, larger offsets reduce LNA linearity until partially rejected and brought within range.



    

\section{Conclusion} \label{sec:conclusion}




A PVT-tolerant low-power mixed-signal AFE with autonomous DC offset
cancellation has been presented, integrating an on-chip fixed-point IIR
filter and bulk-voltage control of the LNA input pair. It achieves
40.55\,dB gain, 3.9\,\textmu V$_\mathrm{rms}$ typical LFP-band noise and 11.11 \,\textmu V$_\mathrm{rms}$ typical EAP-band noise. Finally, it achieves
$\pm$58.13\,mV offset rejection across the bandwidth of operation, suited to
scalable multi-channel implants.

\vfill\null

\bibliographystyle{IEEEtran}
\bibliography{IEEEabrv,Section/references}

\end{document}